\documentstyle[12pt]{article}
\newcommand{\Bs}{B^0_s}
\newcommand{\B}{B^0}

\newcommand{\BB}{\bar B^0}

\newcommand{\be}{\begin{equation}}
\newcommand{\ee}{\end{equation}}

\newcommand{\bea}{\begin{eqnarray}}
\newcommand{\eea}{\end{eqnarray}}
\newcommand{\ba}{\begin{array}}
\newcommand{\ea}{\end{array}}
\newcommand{\ovl}{\overline}
\newcommand{\noi}{\noindent}
\newcommand{\beqa}{\begin{eqnarray}}
\newcommand{\eeqa}{\end{eqnarray}}

\begin{document}
\def\lsim{\;\raise0.3ex\hbox{$<$\kern-0.75em\raise-1.1ex\hbox{$\sim$}}\;}
\def\gsim{\raise0.3ex\hbox{$>$\kern-0.75em\raise-1.1ex\hbox{$\sim$}}}
\begin{titlepage}
\hspace*{11.2cm}HU-TFT-94-30

\hspace*{10.9cm} \makebox[3.0cm]{(hep-ph/yymmnn)}
\vskip 4.5 cm
\begin{center}
{\bf PHENOMENOLOGICAL DETERMINATION OF THE BEAUTY MESON DECAY PARAMETER $f_B$
AND THE CP-VIOLATING ANGLE $\delta$}
\end{center}
\vskip 1.0 cm
\begin{center}
{\bf J. Maalampi}\\
Department of Theoretical Physics,\\
 University of Helsinki, Helsinki, Finland \\
 and \\
{\bf M. Roos}\\ High Energy Physics Laboratory, \\
University of Helsinki, Helsinki, Finland
\vskip 2.0 cm
{\bf Abstract}
\end{center}

We fit the ${\cal CKM}$-matrix to all recent data  with the following free
parameters:  three mixing angles, the CP-violating angle $\delta$ in the Maiani
parametrisation, the top quark mass $m_t$, and the product
$f_B{\cal B}_{\B}^{1/2}$, where $f_B$ is the $B$-meson decay parameter and
${\cal B}_{\B}$ is the bag parameter. Our fits span a contiguous region in the
$(f_B{\cal B}_{\B}^{1/2},\ \cos\delta)$--plane, limited by
$0.117\lsim f_B{\cal B}_{\B}^{1/2}/{\rm GeV}\lsim 0.231$ and --0.95 $\lsim$
$\cos\delta$ $\lsim$ 0.70. The
parameters $f_B{\cal B}_{\B}^{1/2}$ and $\cos\delta$ are strongly  positively
correlated.
\end{titlepage}

When $\B-\BB$ mixing was first discovered \cite{argus} this offered a way
to estimate the mass of the top quark, since the box amplitude responsible for
mixing is dominated by the top exchange. The $\B-\BB$ mixing actually
determines only the product  $f_B{\cal B}_{\B}^{1/2}m_t^2F(m_t^2)$, where
$f_B$ is the unknown pseudoscalar decay constant of the $\B$,  ${\cal B}_{\B}$
is the unknown bag parameter, and $F(m_t^2)$ is a known smooth function of the
top mass (given below). To estimate $m_t$ from this product some theoretical
input \cite{Shifman} of the QCD
quantity  $f_B{\cal B}_{\B}^{1/2}$ was needed.  The mixing data combined with
other charged current input then predicted a top mass value $m_t\gsim 100$ GeV
\cite{MR},\cite{altarelli},\cite{schubert}.

The situation has now changed when the first experimental determination of
$m_t$ is available.  This allows one to reverse the problem, using the charged
current data and $m_t$ to obtain a phenomenological and less model-dependent
estimate of  $f_B{\cal B}_{\B}^{1/2}$.  This information may then be confronted
with the predictions of theoretical models and lattice calculations, which
differ quite substantially from each other, ranging from about 115 MeV
\cite{Shifman} to 300 MeV \cite{Allton}.

In this paper we update our previous analysis of the ${\cal CKM}$ matrix
\cite{MR} by taking into account the new CDF result for $m_t$ \cite{CDF}, and
by using the most recent data for other relevant observables: the ${\cal CKM}$
matrix elements from various charged current processes, the CP-violation
parameter $\vert\epsilon\vert$ from the neutral kaon system, and the $\B-\BB$
mixing parameters $\chi_{d,s}$. The input data are collected in Table 1. For
$m_t$ we have used a value obtained by combining the CDF result
$m_t=174\pm 10^{+13}_{-12}$ GeV \cite{CDF}
with the indirect value  from the fits of the LEP data, the deep inelastic
neutrino-nucleon scattering data and the W mass measurement, $m_t=164^{+16\
+13}
_{-17\ -21}$ GeV \cite{Hol}.

The formalism used to relate the ${\cal CKM}$ matrix elements to various
experimental input is well known ($cf.\ \ e.g.$ \cite{MR,Ali}).  What is new in
our approach compared with our previous analysis \cite{MR} is a quite trivial
change in the CP-violating parameter: we use the cosine of the angle  $\delta$
(in the Maiani parametrisation) rather than the sine.  This choice has the
advantage of exhibiting explicitly that the allowed region in the
$(f_B{\cal B}_{\B}^{1/2},
\cos\delta)$-space is contiguous, in contrast to previous analyses which found
separate solutions in the first and the second quadrant of $\delta$.

 The quantities measured in $\B-\BB$ mixing experiments are the probability
fractions
\be
\chi_{d,s}=\frac{P(B_{d,s}\to \overline B_{d,s})}
{P(B_{d,s}\to  B_{d,s}) + P(B_{d,s}\to \overline B_{d,s})},
\ee
which can be expressed in the form $\chi_{d,s}=x_{d,s}^2(1+x_{d,s}^2)^{-1/2}$,
where
\be
x_q=\frac{G_F^2m_W^2}{6\pi^2}\tau_{B_q}m_{B_q}
(f_{B_q}^2{\cal B}_{B_q})\eta_{B}\frac{m_t^2}{m_W^2}
F(\frac{m_t^2}{m_W^2})\vert V^*_{tq}V_{tb}\vert^2\ \ \ (q=d,s).
\label{x}\ee
Here the function F is defined by \cite{Inami}
\be
F(x)=\frac{1}{4} +
\frac{9}{4}\frac{1}{1-x}-\frac{3}{2}\frac{1}{(1-x)^2}-\frac{3}{2}\frac{x^2\ln
x}{(1-x)^3}.
\ee

\noi
The hard QCD correction factor $\eta_{B_q}$ depends quite strongly on the top
mass. For $m_t=174 $ GeV one finds from \cite{BurasJamin} $\eta_B\simeq 0.49$.
The experimental averages for the $B_{d}^0$ and $B_{s}^0$ lifetimes are
\cite{Danilov}
\beqa
\tau_{B_d^0}&=&1.48\pm 0.10\ {\rm ps},\nonumber\\
\tau_{B_s^0}&=&1.26_{-0.17}^{+0.22}\ {\rm ps}.
\eeqa

\noi
For the masses of the neutral beauty mesons we will use \cite{Danilov,PDG}
\beqa
m_{B_d}&=&5.2790\pm 0.0020\ {\rm GeV},\nonumber\\
m_{B_s}&=&5.3732\pm 0.0042\ {\rm GeV}.
\eeqa

The only unknown parameters, apart from the ${\cal CKM}$ matrix elements, in
(\ref{x}) are the decay constant $f_{B_q}$ and the bag parameter $B_{B_q}$.

Combining the most recent results of ARGUS and CLEO \cite{Ven} gives for
$\chi_d$ the value
\be
\chi_d=0.152\pm 0.030.
\ee
\noi
In the experiments where both $B_d\ovl B_d$ and $B_s\ovl B_s$
are produced one measures the  sum $\chi=f_d\chi_d+f_s\chi_s$
where $f_d$ and $f_s$ are the abundances of $B_d$ and $B_s$ in the b-quark jet.
A recent average of all existing results is given by Danilov \cite{Danilov}:
\be
\chi=0.121\pm 0.010.
\ee
\noi
In applying this quantity we use  $f_d=0.375$ and $f_s=0.15$.

\noi
Furthermore,  we shall assume a fixed SU(3) breaking ratio from lattice
calculations \cite{Abada}
\begin{eqnarray}
f_{\Bs}^2{\cal B}_{\Bs}/f_{\B}^2{\cal B}_{\B}=1.19\ ,
\end{eqnarray}
to which the fit is quite insensitive.

The theoretical expression for the CP-violation parameter $\epsilon$ depends on
the poorly known bag factor ${\cal B}_K$, whereas $\epsilon$ itself is
well measured ($\vert\epsilon\vert=(2.26\pm 0.02)\times10^{-3}$ \cite{PDG}).
Thus the proper  procedure is to use a constraint for ${\cal B}_K$ expressed in
terms of  a constant $\epsilon$ without error. For ${\cal B}_K$ we use the
value
 \beqa
{\cal B}_K= 0.73\pm0.13\ ,
\label{BK}\eeqa

\noi
which covers the values
from different lattice and $1/N$ expansion evaluations
quoted by ref. \cite{Mac}.
The theoretical expression for $\vert\epsilon\vert$ is obtained essentially as
the imaginary part of the box amplitude for the neutral kaon mixing
\cite{BurasSS}.

Experimental results \cite{NA31,E731} on the parameter $\vert\epsilon'\vert $,
describing CP violation in $K^0\to\pi\pi$ decays, are  controversial,
and the experimental accuracy of this quantity is poor. Moreover, some terms in
the theoretical expression are still imprecisely known. We shall therefore not
include $\vert\epsilon'\vert$ in our analysis.

There are 7 free parameters and 14 constraints in the fit: the three mixing
angles  in the ${\cal CKM}$ matrix, the CP-phase $\delta$, the  parameter
$f_B{\cal B}_{\B}^{1/2}$, and the top quark mass $m_t$ (the seventh parameter
is a quark ratio $\kappa$ entering two constraints on $|V_{cs}|$).  The
constraints can be fitted excellently with $\chi^2=8.6$ for 7 degrees of
freedom. There is obviously no reason to increase the "theoretical errors"
further on various input  parameters, as some people would advocate, because
that would just make the fit too good.

The conventional definition of errors on the parameters is always to increment
the best fit $\chi^2$ by 1. However, since we are mainly interested in the
simultaneous 68.3 \% confidence region for the two parameters $f_B{\cal
B}_{\B}^{1/2}$ and $\cos\delta$,  one should increment the best fit $\chi^2$ by
2.3.  This then gives the contour in Fig.~1. The errors on the mixing angles
are of less interest, these parameters being  rather unphysical, and the fit
error on $m_t$ is essentially  equal to the experimental error.

Our best fit yields the parameters and conventional 1$\sigma$ errors
\begin{eqnarray}
&\sin\theta_{12}=0.2203\pm 0.0008,\nonumber\\
&\sin\theta_{23}=0.048\pm 0.003,\nonumber\\
&\sin\theta_{13}=0.0046\pm 0.0008,\nonumber\\
&m_t=174.5\pm 12.5\ {\rm GeV},\\
&f_{\B}{\cal B}_{\B}^{1/2}=149_{-21}^{+23}\ {\rm MeV},\nonumber\\
&\cos\delta =-0.46_{-0.35}^{+0.55}.\nonumber
\end{eqnarray}

\noi
The elements of the ${\cal CKM}$ matrix obtain the following values:
\begin{eqnarray}
V=\left(\matrix{0.9754&0.2202&-0.0021+0.0041\ i\cr
-0.2199+0.0002\ i&0.9743-0.0000\ i&0.0480\cr
0.0126+0.0040\ i&-0.0464+0.0009\ i&0.9988}\right)
\end{eqnarray}

\bigskip
Fig.~1 shows that $f_{\B}{\cal B}_{\B}^{1/2}$ and $\cos\delta$ are very
strongly positively correlated, and still poorly determined. The $1\sigma$
contour in the  $(f_B{\cal B}_{\B}^{1/2},\ \cos\delta)$--plane, is limited to
0.117 $\lsim f_B{\cal B}_{\B}^{1/2}/{\rm GeV}\lsim$ 0.231 and --0.95 $\lsim$
$\cos\delta$ $\lsim$ 0.70.

Recent lattice calculations of $f_{\B}$ yield the
values $180\pm 50$ MeV \cite{Alexandrou}, $187\pm 37$ MeV \cite{Bernard}
and $200\pm 40$ MeV \cite{Maiani}.
This is in good agreement with the contour in Fig.~1. Some earlier estimates,
e.g. \cite{Allton}, with higher values of $f_ B$ are  instead excluded. Perhaps
one could expect the next improvement in precision to come from the
lattice
calculations; this would then permit to determine $\cos\delta$ well.

On the other hand, if the situation regarding
$\vert\epsilon'\vert/\vert\epsilon\vert$ were clarified theoretically and
well measured experimentally, that would help to pin down $\cos\delta$, and
in consequence $f_{\B}{\cal B}_{\B}^{1/2}$ could be determined more precisely.

\vskip 2cm
\noi
{\bf Acknowledgement.}  This work has been supported by the Finnish Academy of
Science.

\setlength{\itemsep}{5mm}

\vfil\eject

\noi{\bf TABLE CAPTION}

\vskip 2cm

\noi{\bf FIGURE CAPTION}

\noi{\bf Figure 1.} The 68.3 \% confidence level contour for  $f_B{\cal
B}_{\B}^
{1/2}$
and $\cos\delta$ and the best fit point.

\noi{\bf Table 1.} The experimental data used in the analysis and their best
fit
 values.
\vfil\eject
\begin{center}
\begin{tabular}{|llll|}\hline
Quantity & Value & Best fit & Reference \\ \hline
$|V_{ud}|$ & 0.9753$\pm$0.0002 & 0.9754 & \cite{Ras} \\
$|V_{us}|$ & 0.2188$\pm$0.0016 & 0.2202 & \cite{Lop} \\
$|V_{cd}|$ & 0.202$\pm$0.010$^{1)}$ & 0.2199 & \cite{Sha,CCFR,FMMF} \\
$|V_{cs}|$ & 1.07$\pm$0.14$^{1)}$ & 0.9743 & \cite{CDHS,TPS} \\
$|V_{cb}|$ & 0.041$\pm$0.006 & 0.0482 & \cite{Bes} \\
$|V_{cd}/V_{cs}|^2$ & 0.057$\pm$0.016$^{2)}$ & 0.0509 & \cite{MARK3} \\
$|V_{ub}|/|V_{cb}|$ & 0.080$\pm$0.025$^{1,3)}$ & 0.0961 & \cite{Bes,CLEO2} \\
$\kappa |V_{cs}|^2$ & 0.43$\pm$0.06$^{1)}$ & 0.4471 & \cite{Sha,CCFR,FMMF} \\
$\kappa |V_{cs}|^2/|V_{cd}|^2$ &
9.6$\pm$1.2$^{1)}$&9.247&\cite{Sha,CCFR,FMMF}\\
$|V_{ts}|/|V_{cb}|$ & 1.09$\pm$0.36 & 0.9659 & \cite{Ali} \\
$\chi$ & 0.121$\pm$0.010 & 0.1291 & \cite{Danilov} \\
$\chi_d$ & 0.152$\pm$0.030 & 0.1273 & \cite{Ven} \\
${\cal B}_K$ & 0.73$\pm 0.13^{3)}$ & 0.7552 & \cite{Mac} \\
$m_t$ & 171$\pm$13 GeV$^{1)}$ & 174.6 GeV & \cite{Hol,CDF} \\ \hline
\end{tabular}
\end{center}
\vskip 0.5 cm

1) Our average.

2) Asymmetric error; only the fit-side error is used.

3) Includes error due to spread of different theoretical models.

\vspace{2.5cm}

\begin{center}
{\bf TABLE 1.}
\end{center}

\end{document}